\documentclass[aps,pra,twocolumn]{revtex4}
\usepackage{graphicx}
\usepackage{dcolumn}
\usepackage{amsmath}
\usepackage{amssymb}
\usepackage{float}
\usepackage{epstopdf}

\def\itTh{{\mathit{\Theta}}}
\def\itOm{{\mathit{\Omega}}}
\def\itPh{{\mathit{\Phi}}}
\def\rel{_{\rm rel}}

\def\inter{_{\rm int}}
\def\ext{_{\rm ext}}

\def\ft{\tilde{f}}
\def\upz{}
\def\upo{^{(1)}}
\def\upt{^{(2)}}

\def\unt{_{\rm uty}}

\def\dir{_{\rm dir}}
\def\ex{_{\rm ex}}
\def\cor{_{\rm c}}

\def\x{_{\rm x}}
\def\c{_{\rm c}}

\def\eps{\varepsilon}
\def\rv{{\bf r}}

\def\xv{{\bf x}}

\def\kv{{\bf k}}
\def\qv{{\bf q}}

\def\beq{\begin{equation}}
\def\eeq{\end{equation}}

\begin{document}
\title{The Fermionic Density-functional at Feshbach Resonance}
\author{Michael Seidl}
\affiliation{Institute of Theoretical Physics,
University of Regensburg, D-93040 Regensburg, Germany}
\author{Rajat K. Bhaduri}
\affiliation{Department of Physics and Astronomy,
McMaster University, Hamilton, Canada L8S 4M1}
\date{\today}

\begin{abstract}

We consider a dilute gas of neutral unpolarized fermionic atoms at
zero temperature. The atoms interact via a short-range (tunable) attractive
interaction. We demonstrate analytically a curious property of the gas 
at unitarity. Namely, the correlation energy
of the gas, evaluated by second order perturbation theory, has the
same density dependence as the first order exchange energy, and the
two almost exactly cancel each other at Feshbach resonance irrespective of the
shape of the potential, provided $(\mu r_s)\gg 1$. Here $(\mu)^{-1}$ is 
the range of the two-body potential, and $r_s$ is defined through 
the number density, $n=3/(4\pi r_s^3)$. The implications of this
result for universality is discussed. 

\end{abstract}

\maketitle
\section{Introduction}

Consider a dilute gas of $N\gg1$ neutral fermionic atoms
(mass $M$) at $T=0$ interacting with a short-range attractive
potential. In general, the
properties of the dilute gas are determined by the number density $n$,
and the scattering length $a$. The
Hamiltonian of this $N$-particle system reads
\beq
\hat{H}=-\frac{\hbar^2}{2M}\sum_{i=1}^{N}\nabla_i^2
+\sum_{i<j}v\Big(|\rv_i-\rv_j|\Big)~.
\label{Ham}
\eeq
Not written explicitly here, there is also an external potential
$v\ext(\rv)$ that forces the $N$ atoms to stay within a large box with
volume $\itOm$ [where $v\ext(\rv)\equiv0$]. The attractive interaction
potential is assumed to have the 2-parameter form 
\beq
v(r)=-v_0f(\mu r)
\label{vint}
\eeq
where $v_0>0$ is the strength of the interaction, $R_0=\frac1{\mu}$ is
its range, and $f(x)$ is a dimensionless function.

In the true ground state of the Hamiltonian \eqref{Ham} the attractive
atoms may form dimers or even clusters. We are, however, looking for a 
metastable  state where there is a
dilute gas of separated atoms with uniform density $n$, 
satisfying the condition $(\mu r_s)\gg1$ where    
$n=\frac{N}{\itOm}=\frac3{4\pi r_s^3}$. Even then, for a weak $v_0$,
there will be BCS-type pairing, followed by dimer formation as the
strength of the interaction increases. This was predicted long back by 
Leggett~\cite{leggett}, and has been observed experimentally~\cite{regal}.  
For the density functional analysis of the uniform gas at Feshbach
resonance, we shall disregard the BCS condensed pairs in this paper. 

To study the effect of the attractive interaction $v(r)$, we consider
the corresponding atom-atom scattering problem in the relative
s-state. 
Separating the center-of-mass motion, we are left with the relative Hamiltonian
\beq
\hat{H}\rel=-\frac{\hbar^2}{M}\left(\frac{d^2}{dr^2}
+\frac{2}{r}\frac{d}{dr}\right)\,-\,v_0f(\mu r)~.
\label{Hrel}
\eeq
Keeping the range of the
potential small enough such that $(\mu r_s)\gg1$, the strength $v_0$ is
adjusted such that the potential can support a single bound state at
zero energy. This happens when the scattering length 
$a\to\infty$, leaving no length scale from the interaction.
Such a tuning of the interaction is possible experimentally, and gives
rise to Feshbach resonance~\cite{feshbach}. The scattering cross
section in the given partial wave (s-wave in our case) reaches the
unitary limit, and the gas is said to be at unitarity. 
It is then expected to display universal behavior~\cite{baker}. 
Note that at Feshbach resonance, there is no length scale left other
than the inverse of the
Fermi wave number $k_F$, where $k_F=(3\pi^2 n)^{1/3}$.
The energy per particle, $E/N$, as a function of the density $n$,
should therefore scale the same way as the
noninteracting kinetic energy, 
$\frac35\hbar^2k_F^2/2M\propto n^{2/3}$. There has been much interest
amongst theorists to calculate the properties of the gas in the unitary
regime ($k_F|a|\gg1$). In particular, at $T=0$, the energy per particle of the
gas is calculated to be 
\begin{equation}
\frac{E}{N}=\xi~\frac{3}{5} \frac{\hbar^2k_F^2}{2M},
\label{free}
\end{equation} 
where $\xi\simeq0.44$~\cite{carlson}. The experimental
value of $\xi$ is about 0.5, but with large error bars~\cite{bart}. Recently,
there have been two Monte Carlo (MC) finite temperature
calculations~\cite{bulag,burovski} of an untrapped gas at unitarity,
where various thermodynamic properties as a function of temperature have
been computed. 
It is clear that at unitarity, the kinetic and
potential energies should scale the same way. This has been assumed 
{\it a priori} in a previous density functional treatment of a 
unitary gas~\cite{pap}. 
However, such a scaling
behavior is not evident from the density functionals for the direct, exchange
and correlation energies \cite{DFT} (see sects. II and III). The aim of the
present paper is to examine this point in some detail. In particular,
we are able to show analytically that the leading contribution of the
correlation energy (calculated in second order perturbation theory), 
cancels the first order exchange enrgy almost exactly at Feshbach
resonance. This happens irrespective of the shape of the potential as
specified by $f(\mu r)$, provided the condition $(\mu r_s)\gg1$. 
We show that our general Eq.(\ref{cancel}) (derived later in the text)
that ensures such a cancellation is satisfied at unitarity 
for a variety of 2-parameter potentials, including the 
square well and the delta-shell, as well as the smoothly varying 
cosh$^{-2}(\mu r)$  and Gaussian potentials. 
This is the main result of the present work. The implications of
this result for universality is marginal. This is because these
potential energy terms, in the limit of $(\mu r_s)\gg 1$, are very
small compared to the kinetic energy ~\cite{baker}. For a moderately 
large value like $(\mu r_s)\simeq 3$ howevr, these terms are
comparable in magnitude to the kinetic energy (sect. IV). Even then, 
the cancellation of the first order exchange, and the second order
perturbative terms leave the direct first order term in tact. In the
electron gas, this (repulsive) term got cancelled by the interaction
of the electrons with the positive ionic background. There is no such
mechanism of cancellation here, unless we assume, rather arbitrarily, 
that the short-range interatomic repulsion cancels this direct
(attractive) contribution.  Even without any such assumptions,
however, our main result (Table I), applicable at Feshbach
reonance,  is interesting from the angle of potential theory.


\section{Perturbation Expansion}

Treating the interaction \eqref{vint} as a weak perturbation in the
Hamiltonian \eqref{Ham}, the unperturbed energy $E^{(0)}$ is the kinetic
energy of a non-interacting Fermi gas,
\beq
E^{(0)}=Nt_s(r_s)\equiv N\frac35\frac{\hbar^2k_F^2}{2M}.
\label{ETF}
\eeq
Here, $k_F=\frac1{\alpha r_s}$ and $\alpha^3=\frac4{9\pi}$.
The corresponding ground state $|\itPh_0\upz\rangle$ is a Slater
determinant of plane waves.

In terms of dimensionless coordinates $\xv_i=\mu\rv_i$, the
Hamiltonian \eqref{Ham} can be written as
\beq
\frac{M}{\hbar^2\mu^2}\hat{H}=
-\frac12\sum_{i=1}^N\nabla_i^2-\lambda\sum_{i<j}f\Big(|\xv_i-\xv_j|\Big),
\quad   \lambda=\frac{Mv_0}{\hbar^2\mu^2}.
\label{Hdim}
\eeq
This suggests that the perturbation parameter is not really small at unitarity.
For example, for the square-well potential, the zero-energy single bound state
occurs when $\lambda=\frac{\pi^2}4$ (see sect.~III).
Nevertheless the low-order terms can point to important information,
even when the expansion is divergent \cite{ISI}.
In our problem, there are three parameters, $\mu $, $ v_0$, and $r_s$. The
unitarity condition relates $\mu$ and $v_0$, so two independent
parameters are left. One of these may be taken to be the small parameter 
$\zeta=(\mu r_s)^{-1}$. The remaining free parameter $v_0$ may be chosen 
independently of $\zeta$ to fulfill the
unitarity condition. 

\subsection{First order}

Formally, the first-order correction,
\beq
E\upo=\langle\itPh_0\upz|\hat{V}\inter|\itPh_0\upz\rangle,
\eeq
has a direct contribution $U(r_s,\mu)=Nu(r_s,\mu)$ with
\begin{eqnarray}
u(r_s,\mu)&=&\frac{\rho^2}{2N}\int_{\itOm}d^3r\int_{\itOm}d^3r'v\Big(|\rv-\rv'|\Big)\nonumber\\
          &=&-\frac32\frac{v_0}{(\mu r_s)^3}f_2.
\label{Udir}
\end{eqnarray}
Here, $f_2=\int_0^{\infty}dx x^2 f(x)$. 

The other first-order contribution is the exchange energy
$E\x(r_s,\mu)=Ne\x(r_s,\mu)$, \cite{baker}
\beq
e\x(r_s,\mu)=-\frac{3k_F}{\pi}\int_0^{\infty}dr j_1(k_Fr)^2 v(r).
\eeq
Here, $j_1(z)$ is a spherical Bessel function. Since $v(r)$ is
short-range and $k_F$ is small in a dilute gas, we can use the
small-$z$ expansion $j_1(z)=\frac{z}3+O(z^3)$ to find
\beq
e\x(r_s,\mu)=\frac34\frac{v_0}{(\mu r_s)^3}f_2+O(\mu r_s)^{-5}.
\label{ex}
\eeq
\subsection{Second order}

\subsubsection{General expressions}

Also the second-order correction,
\beq
E\upt=-\sum_{n\neq0}
\frac{|\langle\itPh_n\upz|\hat{V}\inter|\itPh_0\upz\rangle|^2}{E\upz_n-E\upz_0}
=N\Big[e\upt\dir+e\upt\ex\Big],
\eeq
has a direct and an exchange contribution \cite{GMB},
\begin{eqnarray}
e\upt\dir(r_s,\mu)=-\frac3{32\pi^5}\left(\frac{2M}{\hbar^2\mu^2}\right)v_0^2
\frac{k_F^4}{\mu^4}\int d^3q\,\ft\!\left(\frac{k_F}{\mu}q\right)^2\times\nonumber\\
\times\int_D\frac{d^3k_1\,d^3k_2}{\qv\cdot(\qv+\kv_1-\kv_2)},\hspace*{1cm}
\label{e2dir}
\end{eqnarray}
\begin{eqnarray}
e\upt\ex(r_s,\mu)=+\frac3{64\pi^5}\left(\frac{2M}{\hbar^2\mu^2}\right)v_0^2
\frac{k_F^4}{\mu^4}\int d^3q\,
\ft\!\left(\frac{k_F}{\mu}q\right)\times\nonumber\\
\times\int_Dd^3k_1\,d^3k_2\frac{\ft\!\left(\frac{k_F}{\mu}|\qv+\kv_1-\kv_2|\right)}
{\qv\cdot(\qv+\kv_1-\kv_2)}.\hspace*{1cm}
\label{e2ex}
\end{eqnarray}
While $v_0^2(2M/\hbar^2\mu^2)$ has the dimension energy, the integration
variables $\qv$, $\kv_1$, and $\kv_2$ are dimensionless here. The domain
of the integral over $d^3k_1\,d^3k_2$ depends on $\qv$,
\beq
D:\quad |\kv_1|,|\kv_2|<1;\quad |\kv_1+\qv|,|\kv_2-\qv|>1.
\eeq
Furthermore, $\ft(y)$ is a dimensionless transform of $f(x)$,
\begin{eqnarray}
\ft(y)&=&\int_0^{\infty}\!\!\!dx\,x^2\,f(x)\,j_0(yx)\nonumber\\
&\equiv&\frac1y\int_0^{\infty}\!\!\!dx\,x\,f(x)\,\sin(yx).
\label{ft}
\end{eqnarray}
To recover Eqs.~(8) and (9) of Ref.~\cite{GMB}, put $M=m_e$,
$v_0=-e^2\mu$, and $f(x)=\frac1x$ or $\ft(y)=\frac1{y^2}$, such
that $v(r)=\frac{e^2}r$ becomes the electronic Coulomb repulsion. (Note
that Ref.~\cite{GMB} uses Rydberg units, $m_ee^4/2\hbar^2=e^2/2a_B=1$.)

\subsubsection{The limit $\mu r_s\gg1$}

For a dilute gas (small $k_F$) with short-range interaction (large $\mu$),
Eqs.~\eqref{e2dir} and \eqref{e2ex} can be evaluated in the limit
$\mu/k_F\equiv\alpha\mu r_s\gg1$ where $\alpha^3=\frac4{9\pi}$.
Following Ref.~\cite{Zecca}, we choose a number $q_1$ such that
$1\ll q_1\ll \mu/k_F$ and split the integrals over $d^3q$ into two parts,
\beq
\int d^3q=\int_{q<q_1}\!\!\!d^3q+\int_{q>q_1}\!\!\!d^3q.
\label{split}
\eeq
In the first part with $q<q_1$, we have $q\ll\mu/k_F$ and
$|\qv+\kv_1-\kv_2|\ll\mu/k_F$ (note that $|\kv_1|,|\kv_2|<1\ll q_1$).
Therefore, we may expand $\ft(y)=f_2+O(y^2)$ in Eqs.~\eqref{e2dir} and
\eqref{e2ex} and keep the leading term $f_2$ only. The sum of the two
resulting $q<q_1$ contributions reads
\begin{eqnarray}
e_{q<q_1}\upt(r_s,\mu)=
-\frac3{64\pi^5}\left(\frac{2M}{\hbar^2\mu^2}\right)
v_0^2\frac{k_F^4}{\mu^4}f_2^2\times
\nonumber\\
\times\int_{q<q_1}\!\!\!\!\!\!d^3q
\int_D\frac{d^3k_1\,d^3k_2}{\qv\cdot(\qv+\kv_1-\kv_2)}.\hspace*{1cm}
\label{ec2s}
\end{eqnarray}
The number $q_1$ can be chosen independently of $\mu/k_F\gg1$, despite the
condition $1\ll q_1\ll \mu/k_F$.
Then, the integral in Eq.~\eqref{ec2s} is a finite
constant and we conclude \cite{Zecca}
\beq
e_{q<q_1}\upt(r_s,\mu)=O(\mu r_s)^{-4}.
\label{ec2so}
\eeq 
In the second part $q>q_1\gg1$ of the integral \eqref{split}, we can put
$\qv+\kv_1-\kv_2\approx \qv$, since $|\kv_1|,|\kv_2|<1$. The resulting
contributions to Eqs.~\eqref{e2dir} and \eqref{e2ex} add up to
\begin{eqnarray}
e_{q>q_1}\upt(r_s,\mu)=
-\frac3{64\pi^5}\left(\frac{4\pi}3\right)^2
\left(\frac{2M}{\hbar^2\mu^2}\right)v_0^2\frac{k_F^4}{\mu^4}\times\nonumber\\
\times\int_{q>q_1}\frac{d^3q}{q^2}\,\ft\!\left(\frac{k_F}{\mu}q\right)^2
\hspace*{1cm}
\label{ec2l}
\end{eqnarray}
where $\int_Dd^3k_1\,d^3k_2=(\frac{4\pi}3)^2$ has been used. Now,
\beq
\int_{q>q_1}\frac{d^3q}{q^2}\,\ft\!\left(\frac{k_F}{\mu}q\right)^2
=\,\frac{\mu}{k_F}\,4\pi\int_{y_1}^{\infty}dy\,\ft(y)^2
\label{inty}
\eeq
where $y_1=k_Fq_1/\mu\ll1$. If $\int_{y_1}^{\infty}dy\,\ft(y)^2$ in
Eq.~\eqref{inty} did not depend on $y_1$, expression \eqref{ec2l} did 
rigorously have the order $O(\mu r_s)^{-3}$. However, using the small-$y$
expansion $\ft(y)=f_2+O(y^2)$, we have $\int_0^{y_1}dy\ft(y)^2=
f_2^2y_1+O(y_1^3)$. Consequently, shifting the lower limit $y_1$ of the
integral \eqref{inty} to zero does not affect the leading-order contribution
to expression \eqref{ec2l},
\beq
e_{q>q_1}\upt(r_s,\mu)=O(\mu r_s)^{-3}.
\label{ec2lo}
\eeq 
Therefore, the quantity \eqref{ec2so} does not contribute to the leading order
of $e\cor\upt=e\upt\dir+e\upt\ex$ which is 
purely due to expression \eqref{ec2l},
\beq
e\cor\upt(r_s,\mu)=-\frac3{4\pi}\left(\frac{2M}{\hbar^2\mu^2}\right)
\frac{v_0^2}{(\mu r_s)^3}F+O(\mu r_s)^{-4}
\label{ec2lo}
\eeq
where $F=\int_0^{\infty}dy\ft(y)^2$.

\section{Density scaling at unitarity}

If the perturbation expansion is convergent \cite{GMB}, the total energy
$E(r_s,\mu)=Ne(r_s,\mu)$ of the gas can be expressed in the form
\beq
e(r_s,\mu)=t_s(r_s)+e\x(r_s,\mu)+\sum_{n=2}^{\infty}e\c^{(n)}(r_s,\mu).
\eeq
At unitarity, when the relative Hamiltonian \eqref{Hrel} has a single
bound state at zero energy, the exchange plus correlation energy
$e\x+\sum_{n=2}^{\infty}e\c^{(n)}$ should display the same density scaling
as the kinetic energy, $t_s(r_s)\propto r_s^{-2}\propto\rho^{2/3}$. This is
obviously not the case with any one of the present (leading-order) results
\eqref{ex} and \eqref{ec2lo}. However, since the exchange energy \eqref{ex}
and the second-order correlation energy \eqref{ec2lo} have opposite signs,
they can cancel each other at some value of $\mu$. This happens when
\beq
\frac{Mv_0}{\hbar^2\mu^2}=\frac{\pi f_2}{2F},
\label{cancel}
\eeq
where $f_2=\int_0^{\infty}dx\,x^2\,f(x)$ and $F=\int_0^{\infty}dy\ft(y)^2$.
This is the main result of our paper, and we check it by taking four
different potentials. The results of this analysis, summarized in Table I,
are discussed in detail below.

Generally, we need an eigenfunction $\psi(r)=\frac{u(r)}r$
of the relative Hamiltonian \eqref{Hrel} with eigenvalue zero.
Writing $u(r)=\phi(\mu r)$, the corresponding dimensionless Schr\"odinger
Equation reads
\beq
\phi''(x)=-\lambda f(x)\phi(x),\qquad\lambda\equiv\frac{Mv_0}{\hbar^2\mu^2}.
\label{seqgen}
\eeq
Precisely, we wish to determine that particular value $\lambda\unt$ of
$\lambda$ for which this zero-energy solution is the only bound state.
Then, $\phi(x)$ must obey $\phi(0)=0$, $\phi'(x)<0$ for $x\geq0$, and
$\phi(x)\to$ const.~for $x\to\infty$.
In the following examples (A-D), the solution $\phi(x)$ can be found
analytically or numerically.

(A) Square-well potential of radius $R_0=1/\mu$:
\beq
v(r)=-v_0\itTh(R_0-r)~,
\label{sqw}
\eeq 
where $\itTh(z)$ denotes the Heavyside step function, $\itTh(z)=1$
for $z>0$ and $\itTh(z)=0$ for $z\leq0$. By setting the dimensionless
variable $\mu r=x$, we see that $f(x)=\itTh(1-x)$. The square-well potential
\eqref{sqw} supports a single zero energy bound state when the LHS of
Eq.(\ref{cancel}) is $\lambda\unt=\pi^2/4$.
It may be easily checked analytically that for the square-well
potential \eqref{sqw},
$f_2=\frac13$ and $F=\frac{\pi}{15}$ so that the RHS of 
Eq.(\ref{cancel}) is $\frac52$, very close to its LHS,  $\pi^2/4=2.47$.

(B) Rosen-Morse hyperbolic potential~\cite{carlson}. This potential
is given by 
\beq
v(r)=-v_0~ \mbox{sech}^2 (\mu r)~,
\eeq
which suppotrs a single zero energy bound state when the LHS of 
Eq.(\ref{cancel}) is $\lambda\unt=2$ instead of $\pi^2/4$.  
For this potential, it is easy to check that $f_2=\pi^2/12$. The
quantity $F$, however, has to be calculated numerically, and is given
by $F=0.596$. Again, Eq.(\ref{cancel}) is approximately satisfied,
since its RHS for this potential is $2.17$.

(C) Delta-shell potential~\cite{gottfried}.
Consider the potential 
\begin{eqnarray}
v(r)&=&-\eta \frac{\hbar^2}{M} \delta (r-R_0)~,\nonumber \\
    &=&-\eta \frac{\hbar^2}{M}\frac{1}{R_0} 
        \delta\left(\frac{r}{R_0}-1\right)~,\nonumber \\
    &=&-v_0 f(\mu r)~.
\end{eqnarray}
Thus, we have $v_0=\eta\frac{\hbar^2}{MR_0}$, $\mu=\frac1{R_0}$, and 
$f(x)=\delta(x-1)$. So we get $f_2=1$, $\tilde{f}(y)=\frac{\sin y}{y}$,
and $F=\frac{\pi}{2}$. Hence the RHS of Eq.~(\ref{cancel}) is unity. The LHS
is  $(\eta R_0)$, which is exactly unity when the s-state scattering length
goes to infinity~\cite{gottfried}. Thus Eq.(\ref{cancel}) is exactly obeyed
in this case.

(D) Gaussian Potential.
\beq
v(r)=-v_0 \exp(-\mu^2 r^2)
\eeq
For this example, $f(x)=\exp(-x^2)$ in Eq.~\eqref{vint}.
We find $f_2=\frac14\sqrt{\pi}$ and $F=\frac18(\frac{\pi}2)^{3/2}$
so that the RHS of Eq.~\eqref{cancel} becomes $\pi f_2/2F=2^{3/2}$.
Solving Eq.~\eqref{seqgen} numerically for this $f(x)$, we obtain a 
single bound state at zero energy when the LHS of Eq. (\ref{cancel}) 
is $\lambda\unt=0.949\times2^{3/2}$, close to $2^{3/2}$.\\

\begin{table}[htb]
\begin{center}
\caption{The moments $f_2$ and $F$ of four different profiles $f(x)$
for the potential \eqref{vint}. $\lambda\unt$ is the value at unitarity
of the parameter $\lambda$ in Eq.~\eqref{seqgen}. At unitarity, the ratio $Q$
of the LHS of Eq.~\eqref{cancel} to the RHS is always close to 1.}
\vspace{2mm}
\label{tab:Vee}
\begin{tabular}{|c|cccr|}\hline
\rule[-3mm]{0mm}{8mm} $f(x)$        & $f_2$ & $F$ & $\lambda\unt$ & $Q\quad$ \\ \hline
\rule[-3mm]{0mm}{8mm} $\Theta(1-x)$ & $\frac13$           &
$\frac{\pi}{15}$             & $\frac{\pi^2}4$ & 0.987\; \\
\rule[-3mm]{0mm}{8mm} sech$(x)^2$   & $\frac{\pi^2}{12}$  &
0.596                        & 2               & 0.922\; \\
\rule[-3mm]{0mm}{8mm} $\delta(1-x)$ & 1                   &
$\frac{\pi}2$                & 1               & 1.000\; \\
\rule[-3mm]{0mm}{8mm} exp$(-x^2)$   & $\quad\frac14\sqrt{\pi}$ &
$\quad\frac18(\frac{\pi}2)^{3/2}$ & $\quad$2.684           & $\quad$0.949\; \\ \hline
\end{tabular}
\end{center}
\end{table}

Note, however, that contributions $O(\mu r_s)^{-3}$ may also come from higher
order terms of the perturbation expansion in section II, since that expansion
is carried out with respect to the parameter $\lambda=Mv_0/\hbar^2\mu^2$,
but not $1/\mu r_s$.


\section{Discussion}

The dimensionless Hamiltonian $\hat{h}$ from Eq.~\eqref{Hdim}
depends on the dimensionless paramaters
\beq
\lambda=\frac{Mv_0}{\hbar^2\mu^2}
\label{app1}
\eeq
and, not written explicitly, $x_s=\mu r_s$. The perturbation
expansion of the ground-state energy of $\hat{h}$ reads
\beq
\eps(x_s,\lambda)=\sum_{n=0}^{\infty}\eps_n(x_s)\lambda^n.
\eeq
The ground-state energy of the original Hamiltonian $\hat{H}$,
with three independent parameters, is then given by
\begin{eqnarray}
E(r_s,\mu,\lambda)&=&\frac{\hbar^2\mu^2}{M}\eps(\mu r_s,\lambda)\nonumber\\
&=&\frac{\hbar^2\mu^2}{M}\sum_{n=0}^{\infty}\eps_n(\mu r_s)\lambda^n.
\label{enexp}
\end{eqnarray}
For $\mu r_s\gg1$, we may expand
\beq
\eps_n(\mu r_s)=\sum_{m=0}^{\infty}\frac{\eps_{nm}}{(\mu r_s)^m}.
\label{enexp2}
\eeq
From Eq.~\eqref{ETF}, we have $\eps_{02}=N\frac3{10}\alpha^{-2}$ while
$\eps_{0m}=0$ for $m\neq2$. Eqs.~\eqref{Udir} and \eqref{ex} imply that
$\eps_{1m}=0$ for $m<3$ and $\eps_{13}=N(-\frac32+\frac34)f_2$.
Eventually, due to Eq.~\eqref{ec2lo}, $\eps_{2m}=0$ for $m<3$ and
$\eps_{23}=N(-\frac3{4\pi})2F$.

So far as the unitary point is concerned, we are interested
in a situation where $k_F|a|\gg1\gg k_FR_0\sim(\mu r_s)^{-1}$. In view
of the fact that the perturbation series above does not converge at
unitarity, how significant is our low order perturbation calculation 
in this situation ? 
Note that our first order direct and exchange (potential) energy terms 
given by Eqs. (\ref{Udir},\ref{ex}) are the same as those obtained in 
the Hartree-Fock calculation (see, for example, Eq.(10) of Heiselberg~
\cite{baker}). How big are these terms at unitarity compared to the
kinetic energy per particle ? Taking the example of the square-well
potential discussed earlier, it is straight forward to show that 
our exchange term (\ref{ex}) at Feshbach resonance is 
\beq
e_x(r_s,\mu)=\frac{\pi}{18}\left(\frac{9\pi}{4}\right)^{1/3} \frac{E_F}{\mu r_s}~.
\label{ap1}
\eeq  
For the square-well example, 
\beq
(k_Fa)=\left(\frac{9\pi}{4}\right)^{1/3} \frac{1}{(\mu r_s)} 
\left[1-\frac{\tan \sqrt{\lambda}}{\sqrt{\lambda}}\right]~.
\label{ap3}
\eeq
At unitarity, the RHS diverges for any finite value of $(\mu r_s)$,
how ever large. Even in the neighbourhood of unitarity, it is possible
to have $(k_F|a|) \gg 1$ for $(\mu r_s)\gg 1$. From Eq.(\ref{ap1}),
we note that too large a choice for $(\mu r_s)$ would make $e_x$ negligible
against $\frac35E_F$.  
Instead, taking a modestly large value, $\mu r_s=3$, we obtain the
ration of $e_x$ to kinetic energy per particle to be about
$0.56$. Noting that $e_x$ has a different density-dependence than the
kinetic energy per particle, its cancellation with the second order
perturbative correlation term helps towards scale invariance, but only
if there is a mechanism for the direct first order term to be
cancelled. 

We conclude by emphasizing that the new result in this paper is
displayed in Table 1, and should be of interest from the point of view
of potential theory.  

The authors would like to thank Brandon van Zyl for discussions. This
research was financed by NSERC of Canada.

\newpage


\begin{thebibliography}{99} 
\bibitem{leggett} A.J. Leggett, in {\it Modern Trends in the Theory of
  Condensed Matter}, Springer-Verlag Lecture Notes, Vol. 115, edited by 
  A. Peklaski and J. Przystawa (Springer-Verlag, Berlin, 1980), p.13
\bibitem{regal} C.A. Regal {\it et al.}, Nature (London) {\bf 424}, 47
  (2003); M.W. Zwierlein {\it et al.}, Phys. Rev. Lett. {\bf91}, 250401
  (2003); C.A. Regal {\it et al.}, Phys. Rev. Lett. {\bf 92}, 040403 (2004); 
  M.W. Zwierlein {\it et al.}, Nature (London) {\bf 435}, 1046
  (2005); G. B. Partridge {\it et al.}, Science {\bf 311}, 503 (2006).
\bibitem{feshbach} S. Inouye {\it et al.}, Nature (London) {\bf 392},
  151 (1998); Ph. Courteille {\it et al.}, Phys. Rev. Lett {\bf 81},
  69 (1998).
\bibitem{baker} G.A. Baker, Phys. Rev. {\bf C60}, 054311 (1999); 
  H. Heiselberg, Phys. Rev. {\bf A63}, 043606 (2001);
  T.-L. Ho. Phys. Rev. Lett. {\bf 92}, 090402 (2004).
\bibitem{carlson} J. Carlson, S.-Y. Chang, V. R. Pandharipande, and
  K. E. Schmidt, Phys. Rev. Lett. {\bf 91}, 050401 (2003);
  A. Perali, P. Pieri, and G. C. Strinati, Phys. Rev. Lett. {\bf 93}, 
  100404 (2004).
\bibitem{bart} M. Bartenstein {\it et al.}, Phys. Rev. Lett. {\bf 
92},   120401 (2004); T. Bourdel {\it et al.}, Phys. Rev. Lett. {\bf 
93}, 050401   (2004).
\bibitem{bulag} A. Bulgac, J. E. Drut J.E., and P. Magierski,
  Phys. Rev. Lett. {\bf 96}, 090404 (2006). 
\bibitem{burovski} E. Burovski, N. Prokof'ev,
B. Svistunov, and M. Troyer, Phys. Rev. Lett. {\bf 96}, 160402 (2006).
\bibitem{pap} T. Papenbrock, Phys. Rev. {\bf A72}, 041603 (R) (2005); 
A. Bhattacharyya and T. Papenbrock, Phys. Rev. {\bf A74}, 041602 (R) (2006).

\bibitem{DFT} R. G. Parr and W. Yang, {\it Density-Functional Theory
of Atoms and Molecules} (Oxford University Press, New York, 1989);
W. Kohn, Rev. Mod. Phys. {\bf 71}, 1253 (1999). 
\bibitem{ISI} M. Seidl, J. P. Perdew, and S. Kurth, Phys. Rev. Lett. {\bf 84},
5070 (2000).
\bibitem{GMB} M. Gell-Mann, K. A. Brueckner, Phys. Rev. {\bf 106}, 364 (1957).
\bibitem{Zecca} L. Zecca, P. Gori-Giorgi, S. Moroni, and G. B. Bachelet,
Phys. Rev. B {\bf 70}, 205 127 (2004).
\bibitem{gottfried} K. Gottfried, {\it{Quantum Mechanics}vol.I},
  (W. A. Benjamin, Inc., New York, 1966). See sect. (15).
\end{thebibliography}
\end{document}